\documentclass[prd,tightenlines,nofootinbib,amsfonts,amssymb,amsmath,color,11pt]{revtex4}
\usepackage{graphicx} 

\usepackage{graphicx}
\usepackage{dcolumn}
\usepackage{bm}
\usepackage{epsfig}
\usepackage{color}
\usepackage{slashed}
\usepackage{hyperref}
\hypersetup{
    colorlinks=true,
    linkcolor=red,
    citecolor=blue,
}

\newcommand{\lam}{\lambda}

\newcommand{\cD}{{\cal D}}

\newcommand{\gam}{\gamma}
\newcommand{\del}{\delta}
\newcommand{\ve}{\varepsilon}
\newcommand{\al}{\alpha}
\newcommand{\vphi}{\varphi}

\input epsf

\def\fun#1#2{\lower3.6pt\vbox{\baselineskip0pt\lineskip.9pt
  \ialign{$\mathsurround=0pt#1\hfil##\hfil$\crcr#2\crcr\sim\crcr}}}
\def\simgt{\mathrel{\lower0.6ex\hbox{$\buildrel {\textstyle >}
 \over {\scriptstyle \sim}$}}}
\def\simlt{\mathrel{\lower0.6ex\hbox{$\buildrel {\textstyle <}
 \over {\scriptstyle \sim}$}}}

\def\bea{\begin{eqnarray}}
\def\eea{\end{eqnarray}}
\def\be{\begin{equation}}
\def\ee{\end{equation}}
\def\tr{{\rm tr}\,}
\def\be{\begin{equation}}
\def\ee{\end{equation}}
\def\ba{\begin{eqnarray}}
\def\ea{\end{eqnarray}}
\def\p{\partial}

\def\mc{\mathcal}

\begin{document}

\preprint{}

\title{A Higgs Mechanism for Vector Galileons}

\author{Matthew Hull, Kazuya Koyama and Gianmassimo Tasinato}

\bigskip

\vskip1cm

\affiliation{
\vskip0.3cm
Institute of Cosmology \& Gravitation, University of Portsmouth, Dennis Sciama Building, Portsmouth, PO1 3FX, United Kingdom \\
}


\begin{abstract}
Vector theories with non-linear derivative self-interactions that break gauge symmetries have been shown to have interesting cosmological applications.  In this paper we introduce a way to spontaneously break the gauge symmetry and construct these theories via a Higgs mechanism.  In addition to the purely gauge field interactions, our method generates new ghost-free scalar-vector interactions between the Higgs field and the gauge boson.  We show how these additional terms are found to reduce, in a suitable decoupling limit, to scalar bi-Galileon interactions between the Higgs field and Goldstone bosons.  Our formalism is first developed in the context of abelian symmetry, which allows us to connect with earlier work on the extension of the Proca action.  We then show how this formalism is straightforwardly generalised to generate theories with non-abelian symmetry.  
\end{abstract}
\vspace{1cm}



\maketitle

\section{Introduction}

The nature of dark energy is one of the most profound open problems in  Physics.  Present day cosmic acceleration could be associated with
 some  contribution to the matter  energy momentum tensor,  in addition to matter and radiation. Possible examples are a  positive
  cosmological constant, or some  scalar field whose dynamics make the universe accelerate as in
  scalar-tensor theories; see \cite{Copeland:2006wr} for a   review. 
   Alternatively, this phenomenon can be due to an infrared modification of the gravitational interactions described by 
    Einstein's General Relativity (GR): a recent  review on this topic is given in \cite{Clifton:2011jh}.
     On the other hand, Lovelock's theorem ensures that any consistent modification of the theory of GR plus cosmological constant introduces new degrees of freedom.
 Such theories contain  additional gravitational modes, typically including scalars, that potentially mediate long-range interactions.

This implies that any theory  attempting to explain dark energy 
 (besides a pure cosmological constant)  has to deal
with new fields, whose interactions with matter must be suppressed at sufficiently small scales to satisfy stringent
constraints from the absence of any detectible fifth force. This can be done by screening new interactions by means of either chameleon \cite{Khoury:2003aq} or Vainshtein  \cite{Vainshtein:1972sx} mechanisms. Galileons are a class of scalar-tensor theories that, by exploiting the non-linearity of derivative self-interactions, are able to generate cosmological acceleration while at the same time automatically screen
scalar forces at small scales via a Vainshtein mechanism \cite{Nicolis:2008in}. 
 Interestingly, scalars
with 
Galileon interactions   
can find explicit realizations
as St\"uckelberg fields in theories with 
   broken local symmetries: examples are gravity theories with broken diffeomorphism invariance such as dRGT massive gravity \cite{deRham:2010kj}, or  vector theories with  broken gauge symmetries such as those developed in \cite{Tasinato:2014eka,Heisenberg:2014rta}. Physically, Galileon scalars are associated with the Goldstone bosons from broken continuous symmetries. 
 From a theoretical point of view, the advantage of realizing Galileons is that they come with stringent consistency requirements (in particular the absence of ghosts) and this helps to reduce the size of the parameter space, which makes these set-ups more predictive than generic scalar-tensor scenarios.  Moreover, the Galileonic symmetry underlying these models can play a role in protecting the structure of the theory under radiative corrections.

 So far however, the symmetry in this class of theories has been broken by hand and then recovered at a second stage by adding St\"uckelberg fields with specific interactions.  An issue with applying the St\"uckelberg approach to ensure gauge invariance  is that it is not always easy to de-mix the physical degrees of freedom, especially
  in non-abelian gauge theories,  and thus it can be difficult to verify whether the theory is under perturbative control within the range of interest.  Moreover, unitarity problems can arise:   a typical example  is the scattering amplitude of W-bosons in the Standard Model, which needs to be unitarized by new physics arising below the TeV scale. 
    For these reasons it would be interesting to generalize these constructions, by breaking diffeomorphisms or gauge symmetries  {\it spontaneously} -- for example by a Higgs mechanism-- and yet still be able to recover Galileonic interactions, at least in some limits, for the available degrees of freedom in the broken phase. 
     The advantage of spontaneous breaking is that the underlying symmetry invariance can protect 
     and further restrict 
     the structure of the theory and it can improve the perturbative behavior of scattering amplitudes.
    Moreover, it can additionally provide 
    criteria -- based on symmetry principles -- to extend the abelian theory of \cite{Tasinato:2014eka,Heisenberg:2014rta} to the non-abelian case and possibly determine couplings between the dark energy sector and standard matter, offering new avenues to test the theory.
   A way to spontaneously break diffeomorphism invariance to obtain dRGT massive gravity has yet to be found.  However, we will show that, instead, a Higgs mechanism for vector theories with broken gauge symmetries such as those developed in \cite{Tasinato:2014eka,Heisenberg:2014rta} can be realized.  The gauge symmetry can be spontaneously broken by a Higgs scalar field acquiring a vacuum expectation value, and the theory after symmetry breaking coincides with the broken abelian gauge theory of \cite{Tasinato:2014eka}.
  Additionally, our Higgs mechanism can be straightforwardly extended to scenarios with non-abelian symmetry,
  showing that a Higgs  construction can suggest new ways to straightforwardly generalize the theory of interest to 
  interesting and quite  non-trivial set-ups.

  The Goldstone bosons associated with the broken symmetry are `eaten' by the longitudinal modes of the vector (more precisely, a unitary gauge can be selected that set them to zero).  However, in an appropriate decoupling limit, the dynamics of the vector longitudinal modes correspond to one of the would-be Goldstone bosons which is controlled by Galileon interactions. We show that the interactions of the scalar Higgs field itself also enjoys Galileonic symmetries, and that the Higgs-Goldstone boson system assembles into a specific bi-Galileon combination.

 \smallskip
 
 A Higgs mechanism, by construction, adds some new degrees of freedom to the theory, gauged under the symmetry being considered together with a non-trivial potential that spontaneously breaks this symmetry.  We start in Section \ref{sec-abelian} by discussing the case of abelian interactions. We consider as a fundamental degree of freedom a complex Higgs scalar field charged under the $U(1)$ abelian gauge symmetry, with a classical `Mexican hat' potential.  The new Higgs interactions that we consider correspond to higher dimensional non-renormalizable operators, involving gauge invariant derivative self-couplings of the Higgs field. 
      When the Higgs field
      sits at the minimum of its potential 
      and acquires a vacuum expectation value $v$, the resulting theory corresponds to the vector self-interacting theory studied
       in \cite{Tasinato:2014eka}, with parameters depending on $v$, the gauge coupling constant $g$, as well as 
      on the parameters characterizing the higher derivative Higgs self-interactions.  Moreover, when considering Higgs excitations around its minimum, one finds
      new scalar-vector  derivative interactions -- absent in the original theory that involved vector self-interactions only --
       appearing in  consistent  combinations built in such a way to   avoid the appearance of
         ghosts.
This is    a stringent     requirement that 
 constrains the structure of the  Higgs self-interaction. We determine various examples of  higher dimensional derivative self-interactions
 for the Higgs boson,
  that once expanded around 
  the minimum of the Higgs potential lead to ghost-free derivative interactions between the vector and scalar,
%
 that generalize multi-Galileon constructions to the vector case. 
  We show that in a suitable  decoupling limit the theory 
   reduces to a scalar bi-Galileon theory, that couples with Galileon invariant interactions the Higgs boson with the
   would-be Goldstone modes of the broken symmetry.   In the interest of highlighting the relevance of a Vainshtein-like effect for our model, we conclude the section by briefly discussing a scenario in which the Higgs and the vector are   coupled to external matter.    
    In Section \ref{sec-nonab}  we straightforwardly  extend our constructions to  the case of non-abelian symmetry,
    and discuss some of its physical consequences.

As far as we are aware, this is the first example of a consistent realization of a Higgs mechanism in theories 
with a spontaneously broken symmetry, that lead to Galileonic theories in the remaining degrees of freedom.
  Our set-up can be regarded as a possible step towards finding a consistent UV completion of theories closely related to Galileons.

\section{ Higgs mechanism and  generalized abelian symmetry breaking}\label{sec-abelian}

We discuss a Higgs mechanism that spontaneously breaks an abelian symmetry, in such a way to generate a vector mass term and the class of 
  derivative vector self-interactions studied in \cite{Tasinato:2014eka,Heisenberg:2014rta}.  We work in four dimensional Minkowski space.  It is well known that an abelian symmetry can be broken by a mass term controlled by a scale $m_A$.  However, in addition to this, we can add derivative interactions
for the vector field $A_\mu$, the simplest of which is a dimension-4 operator weighted by a dimensionless coupling, (denoted as $\beta$): 
 \be\label{lav}
 {\cal L}_A\,=\,-m_A^2 A_\mu A^\mu-\beta\, A_\mu A^\mu\,\partial_\rho A^\rho\,.
 \ee
In addition, one can consider a handful of higher-dimensional operators with a similar structure as above. 
These operators break abelian gauge invariance, 
but are nevertheless 
 consistent since the $A_0$-component of the gauge field is a constraint:  its equation of motion does not contain
 time derivatives acting on the field.
   So (\ref{lav}) does not induce ghost degrees of freedom: see \cite{Tasinato:2014eka,Heisenberg:2014rta} for more details. These systems are interesting for 
their cosmological applications and, as we will see, 
 they are related to Galileons, since the dynamics of Goldstone bosons associated with the breaking of symmetry is described by Galilean interactions, at least in an appropriate decoupling limit.

 Interactions as the one in eq.(\ref{lav}) 
can arise by a process of spontaneous breaking of gauge symmetry
via a Higgs mechanism.
Let us consider a gauge invariant action for a complex scalar Higgs field with higher order derivative couplings,
%
%
%
\begin{eqnarray}\label{lagtot}
{\cal L }_{tot} &=& - (\cD_{\mu} \phi)(\cD^{\mu} \phi)^* -
\frac{1}{4} F^{\mu \nu} F_{\mu \nu} - V(\phi) \nonumber \\
&+&
\,\mc{L}_{\mathrm{(8)}}+
\,\mc{L}_{\mathrm{(12)}}+
 \,\mc{L}_{\mathrm{(16)}}\,.
\end{eqnarray}
 The first line contains the usual kinetic terms for scalar and vector ($F_{\mu\nu}=\partial_\mu A_\nu-\partial_\nu A_\mu$) and the Higgs potential.  The 
second line contains  new dimension 8, 12, 16 gauge invariant operators, that are suppressed by a mass scale $\Lambda$, and 
 describe the Higgs derivative self-interactions associated with the pattern of spontaneous symmetry breaking that we are interested in.

The covariant derivative acting on the Higgs field contains the gauge field $A_\mu$, and is defined as
\be
\cD_{\mu} = \p_{\mu} - i g A_{\mu}
\,,
\ee
with $g$ a coupling constant. The Higgs potential has the traditional `Mexican hat' form
\be
V(\phi) = - \mu^2 \phi \phi^* +\frac{\lambda}{2} (\phi \phi^*)^2\,,
\ee
and has a minimum at 
\be
\langle \phi \rangle \equiv v = \left( \frac{\mu^2}{\lambda} \right)^{1/2}\label{vev}\,.
\ee
%

We  demand that
  Lagrangian ${\cal L}_{tot}$ is  invariant under a $U(1)$ gauge symmetry, acting on the scalar and on the vector 
 as
 \bea
 \phi&\to&\phi\,e^{i \,\xi} \,,\\
 A_\mu&\to&A_\mu +\frac{1}{g}\,\partial_\mu \xi 
\,,
 \eea
 for an arbitrary function $\xi$. Under a $U(1)$ transformation, the covariant derivative transforms as
 \bea
 \cD_\mu \phi &\to&e^{i \,\xi}\, \cD_\mu \phi \,,\\
  \cD_\mu \,  \cD_\mu\, \phi &\to&e^{i \,\xi}\, \cD_\mu \,  \cD_\nu\,\phi 
 \,.\eea 
  
  Using the transformation properties of the covariant derivative under gauge transformations, 
   it is straightforward to check that the following
    tensors are gauge invariant:
%
  %
 %
  \begin{align}
	L_{\mu\nu}&\equiv \frac12 \left
	[(\mc{D}_{\mu}\phi)^{*}(\mc{D}_{\nu}\phi)+(\mc{D}_{\nu}\phi)^{*}(\mc{D}_{\mu}\phi)\right]\,,
	 \\ P_{\mu\nu}&\equiv  \frac12 \left[\phi^*\mc{D}_{\mu}\mc{D}_{\nu}\phi +\phi\,\left(
	 \mc{D}_{\mu}\mc{D}_{\nu}\phi\right)^* \right]\,,\\
	  \,Q_{\mu\nu} &\equiv \frac{i}{2} \left[
	  \phi \left( \mc{D}_{\mu}\mc{D}_{\nu}\phi\right)^* 
	  -\phi^*\mc{D}_{\mu}\mc{D}_{\nu}\phi 
	  	  \right]\,. 
\end{align}
Notice that $P_{\mu\nu}$ and $ Q_{\mu\nu}$ are formed by second covariant derivatives: these  contain derivatives
of the vectors,   that are needed to build   derivative vector self-interactions
as in eq.(\ref{lav}). 
Together with the totally antisymmetric $\ve$-tensor in four dimensions (with $\ve_{0123}=1$), the previous tensors are the 
ingredients we use to  
%
define the operators ${\cal L}_{(8),\, (12),\, (16)}$ introduced in the second line of eq.(\ref{lagtot}) as 
  \begin{align} \label{defol}
	\mc{L}_{\mathrm{(8)}} &= 
	\frac{1}{2!\, \Lambda^4}\,\ve^{
	\al\beta
	\mu_{1}\mu_{2}}\ve_{\al \beta\nu_{1}\nu_{2}}\,\left[\, \al_{(8)} L_{\mu_{1}}^{\,\nu_{1}}P_{\mu_{2}}^{\,\nu_{2}}+\beta_{(8)}L_{\mu_{1}}^{\,\nu_{1}}Q_{\mu_{2}}^{\,\nu_{2}}\,\right]&\\
	 \mc{L}_{\mathrm{(12)}} &= \frac{1}{\Lambda^8}\ve^{\al\mu_{1}\mu_{2}\mu_{3}}\ve_{\al\nu_{1}\nu_{2}\nu_{3}}\,\left[\, \al_{(12)} L_{\mu_{1}}^{\,\nu_{1}}P_{\mu_{2}}^{\,\nu_{2}}P_{\mu_{3}}^{\,\nu_{3}}+\beta_{(12)}L_{\mu_{1}}^{\,\nu_{1}}Q_{\mu_{2}}^{\,\nu_{2}}Q_{\mu_{3}}^{\,\nu_{3}}\,\right]&\\\mc{L}_{\mathrm{(16)}} &=\frac{1}{ \Lambda^{12}} \ve^{\mu_{1}\mu_{2}\mu_{3}\mu_{4}}\ve_{\nu_{1}\nu_{2}\nu_{3}\nu_{4}}\,\left[\, 
	\al_{(16)} L_{\mu_{1}}^{\,\nu_{1}}P_{\mu_{2}}^{\,\nu_{2}}P_{\mu_{3}}^{\,\nu_{3}}P_{\mu_{4}}^{\,\nu_{4}} +\beta_{(16)}L_{\mu_{1}}^{\,\nu_{1}}Q_{\mu_{2}}^{\,\nu_{2}}Q_{\mu_{3}}^{\,\nu_{3}}Q_{\mu_{4}}^{\,\nu_{4}}\,\right]&
	 \label{defoq}
\end{align}
that are weighted by dimensionless parameters $\alpha_{(i)}$,  $\beta_{(i)}$, and suppressed by an energy
scale $\Lambda$ to the appropriate powers.  We 
present in Appendix \ref{app-ghostfree} arguments that show that   
    these operators  lead to equations of motion with at most two space-time
 derivatives, analogously to what happens for standard Galileons \cite{Nicolis:2008in}. 
   Indeed, the $\ve$-tensors present in the above definitions have been introduced 
   to automatically avoid the emergence of ghost degrees of freedom. 
 Similar gauge invariant Higgs Lagrangians were also studied in \cite{Kamada:2010qe, Zhou:2011ix}.
   Notice that all these operators are higher-dimensional and hence apparently non-renormalizable: we will return
   to this point at the very end of this section.

  \smallskip

  To understand the physical consequences of these new self-interactions,
   it is convenient  to decompose the complex scalar into its norm and phase: 
\be
\phi\,=\,\varphi\,e^{i g\,\pi}\,,
\ee
where $\varphi$, $\pi$ are two real fields. $\varphi$ does not transform under $U(1)$ 
 gauge symmetry, while the field $\pi$ transforms non-linearly as
  $\pi\,\to\,\pi+\frac{\xi}{g}$: the phase $\pi$ behaves as the would-be Goldstone
 boson for the broken abelian symmetry. 
Hence
defining the gauge invariant  combination
\be\label{ginc}
\hat{A}_\mu\,\equiv\,A_\mu- {\partial_\mu \pi}\,,
\ee
we can express the covariant derivatives as
\bea
{\cal D}_\mu\,\phi&=&\left[\partial_\mu\,\varphi-i\,g\,{\varphi}\,\hat{A}_\mu\right]\,e^{
i g\,\pi
}\,,\label{defcD}
\\
{\cal D}_\mu\,{\cal D}_\nu\,\phi&=&
\left[\partial_\mu\partial_\nu\,\varphi 
-i g \varphi \partial_\mu \hat{A}_\nu-i g \hat{A}_\mu 
\partial_\nu \varphi-i g \hat{A}_\nu 
\partial_\mu \varphi
-g^2 \varphi \hat{A}_\mu \hat{A}_\nu 
\right]\,e^{
{{i g\,\pi}}
}\,,
\eea
with the pieces inside the square parenthesis  invariant under the gauge transformations. It is important
to stress that using this Higgs construction the would-be Goldstone fields combine automatically with the vectors
and appear in the action only 
in the gauge invariant combination (\ref{ginc}).

Using these relations, 
 the operators defined in eqs.(\ref{defol})-(\ref{defoq}) can be expressed as
\begin{align}
	L_{\mu\nu}&= \p_{\mu}\vphi \p_{\nu}\vphi + g^2 \vphi^2 \hat{A}_{\mu}\hat{A}_{\nu}\,,\\
	 P_{\mu\nu}&= \vphi \p_{\mu}\p_{\nu}\vphi - g^2\vphi^2\hat{A}_{\mu}\hat{A}_{\nu} \\Q_{\mu\nu} \,
	  &= \frac{g}{2}\,[\p_{\mu}(\vphi^2\hat{A}_{\nu})+\p_{\nu}(\vphi^2\hat{A}_{\mu})]\,,
\end{align}
which shows that they are symmetric in their two indexes. 
 It is straightforward to plug these expressions into eqs.(\ref{defol}) to derive  explicit forms for the Lagrangians
${\cal L}_{(8),\,(12),\,(16)}$, by also using the following identity involving contractions of the $\ve$-tensors:
\be
\ve_{{\alpha_1}\dots\alpha_{4-n}\alpha_{1}\dots\alpha_n}
\,\ve^{{\alpha_1}\dots\alpha_{4-n}\beta_{1}\dots\beta_n}\,=\,-\left(4-n\right)!\,n!\,
\delta_{\alpha_{1}}^{[\beta_{1}} \ldots \delta_{\alpha_{n}}^{\beta_{n}]}\,.
\ee
where $[\dots]$ denotes weighted index anti-symmetrization. 
For example, let us focus on the lower dimensional interaction contained in ${\cal L}_{(8)}$,
 proportional to the dimensionless coefficient $\beta_{(8)}$. We get


\bea
{\cal L}_{(8)}&=&
-\frac{\beta_{(8)}}{\Lambda^4}\left(L_\rho^{\,\,\rho} Q_\sigma^{\,\,\sigma}-L_{\mu}^{\,\,\nu} Q_{\nu}^{\,\,\mu} \right)
\,,
\\
&=&
-\frac{g\,\beta_{(8)}}{\Lambda^4}
\left( \partial_\mu \varphi
 \partial^\nu \varphi+g^2\,\varphi^2\, \hat A_\mu  \hat A^\nu
\right)\,\partial_\rho (\varphi^2  \hat A^\sigma)\,\left( \delta_\nu^{\,\,\mu}  \delta_\sigma^{\,\,\rho} 
- \delta_\nu^{\,\,\rho}  \delta_\sigma^{\,\,\mu} 
 \right)
\label{expl8}
\,.
\eea

\smallskip

This expression is manifestly gauge invariant, and describes the
 interactions between the norm $\varphi$ of the Higgs field
and the  gauge-invariant combination of vector and  would-be Goldstone bosons.
Additional dimension-8 operators proportional to $\alpha_{(8)}$ 
 could be included, that 
lead to other interactions between   gauge fields
and first derivatives of the scalar $\varphi$; these 
  are of less interest in the present context, so we  
ignore them here.  
Analogous expressions can be straightforwardly obtained for ${\cal L}_{(12)}$, ${\cal L}_{(16)}$: the
resulting formulae are however cumbersome so we will not write them explicitly. We instead move on to discuss
some phenomenological aspects of the Higgs interactions associated with ${\cal L}_{(8)}$.

\bigskip


As we explained, our main motivation   is to generate,
 by the phenomenon of  spontaneous symmetry breaking,     the vector  self-interactions of eq.(\ref{lav}) and their
generalizations discussed in  \cite{Tasinato:2014eka,Heisenberg:2014rta}. 
The phenomenon of spontaneous symmetry breaking is associated with the Higgs developing a vacuum expectation
value  $v$ as in eq.(\ref{vev}),  and acquiring non-trivial dynamics when fluctuating around the minimum of its potential. In order to 
study the dynamics of Higgs fluctuations, it is convenient to expand the norm of the Higgs around the minimum $v$ of the
potential, and write
\be
\varphi=\left(v+\frac{h}{\sqrt{2}}\right)
\ee
which allows us to canonically normalize the Higgs fluctuations $h$. 
By applying this expansion, the initial Lagrangian ${\cal L}_{tot}$ -- including only the $\beta_{(8)}$ contribution 
to ${\cal L}_{(8)}$ written in eq.(\ref{expl8}) -- results
\bea
{\cal L}_{tot}&=&-\frac{1}{4} F_{\mu \nu} F^{\mu \nu}- m_A^2\, \hat{A}^2-\tilde \beta\,\hat A_\mu \hat A^\mu\,\partial_\rho \hat A^\rho
\nonumber
\\
&&- \frac12\,(\p h)^2 -
\frac12\,m_h^2\,
 h^2\,-\frac{\sqrt{\lambda}\,m_h}{2}\,h^3\,-\frac{{\lambda}\,}{{8}}\,h^4
 -\sqrt{2}\,g\,m_A\,h\,\hat{A}_{\mu} \hat{A}^{\mu}-\frac{g^2}{2}\,h^2\,\hat{A}_{\mu} \hat{A}^{\mu}
 \,
\nonumber
\\
&&+\frac{4\,g\,\tilde \beta}{3\,m_A}\,\left(\sqrt{2}\,h +\frac{3\,g}{2\,m_A}\,h^2\, +\frac{g^2}{\sqrt{2}\,m_A^2}\, h^3+\frac{g^3}{8\,m_A^3}\,h^4\right)\,\left(\hat A_\mu\,\hat A^\nu\,\partial_\nu \hat A^\mu-\hat A_\mu\,\hat A^\mu\,\partial_\rho \hat A^\rho \right)
\nonumber
\\
&&+\frac{\tilde \beta }{3\,m_A^2 }\,\left(  1+\frac{\sqrt{2}\,g}{m_A}\,h+\frac{g^2}{2\,m_A^2}\,h^2\right)\left(\partial_\mu  h \,\partial^\nu  h \,\partial_\nu \hat A^\mu-
\partial_\mu  h\, \partial^\mu  h \,\partial_\rho \hat  A^\rho
 \right)\,, \label{explag}
\eea
with
\bea
m_A&=& g \,v\,,
\\
\tilde{\beta}&=&\frac{3 \,g^3\,\beta_{(8)} \,v^4}{2\,\Lambda^4} \label{deftb}
\,,
\\
m_h&=&\sqrt{2\,\lambda}\,v\,,
\eea
where we have neglected the field-independent part of the potential, that contributes to the cosmological 
constant.  

The previous Lagrangian is fully gauge invariant, being expressed in terms of the
gauge invariant combination given in eq.(\ref{ginc}), and describes the dynamics of four degrees of freedom, two scalars
and a massless vector. Choosing the 
 unitary gauge $\pi=0$ enables us 
to analyze the dynamics of the physical degrees of freedom: the Higgs scalar $h$ and a massive gauge boson
$A_\mu$ (again, with a total of four degrees of freedom).
 Working in  the  physically transparent unitary gauge, one finds that the previous Lagrangian eq.(\ref{explag}) leads to several
  interesting interactions.

 In the first two lines one finds renormalizable interactions described by (up to) dimension-4 operators:
  the 
  the Higg's \textit{vev}, $v$, gives a mass to the gauge field, $m_A\,= \,g v$, and
provides the simplest example of a derivative vector self-interaction: that
 of  eq.(\ref{lav}), which was 
studied in \cite{Tasinato:2014eka,Heisenberg:2014rta}. Hence, the phenomenon of
spontaneous symmetry breaking automatically generates  the desired  vector derivative self-interactions;
  the dimensionless coupling constant   $\tilde \beta$ in front of this derivative operator 
   depends on the
 ratio of the Higgs vev $v$ and the scale $\Lambda$,
 see eq.(\ref{deftb}).
 
 On the other hand, we discover that 
 in
addition to these renormalizable derivative vector self-interactions, 
 this  Lagrangian contains new higher dimensional operators between the physical Higgs
field $h$ and the gauge field, contained in the last two lines of  eq.(\ref{explag}). The couplings that govern
 those interactions  
are fixed by the mechanism of symmetry  breaking and gauge invariance, and are suppressed by a mass scale corresponding
to the vector mass $m_A$ to  appropriate powers.  
 Notice that 
 all these new higher dimensional interactions
are derived from our initial  Lagrangian, and consequently are ghost-free since the associated 
equations of motion contain at most two space-time derivatives. It is indeed straightforward to show that for 
all these interactions the $A_0$ component of the gauge field remains a constraint, and the equations of motion
for all the fields contain at most two space-time derivatives (including the new vector-scalar interactions in the last 
line of (\ref{explag})). One can further generalize these results by including the Lagrangians ${\cal L}_{12}$ and  ${\cal L}_{16}$,
that lead to the complete set of derivative vector interactions discussed in \cite{Tasinato:2014eka}, and in addition to new scalar-vector interactions
that generalize the last line of eq.(\ref{explag}), see Appendix \ref{app.B} for details on how they are constructed.
  
\bigskip

It would be very interesting  to study the observational effects of all these new operators: since they are suppressed by powers of $m_A$, they can lead to sizeable effects if $m_A$ is not large. However, screening mechanisms might
occur, similar to what happens with the Vainshtein effect and Galileon interactions  in gravitational set-ups.
The complete phenomenology of the previous system along these lines goes beyond the scope of this work, but let us develop  
some intriguing  relations between the previous system and Galileons.  We return to the fully gauge invariant
Lagrangian (\ref{explag}) before choosing any gauge, with the aim to study the dynamics of would-be Goldstone bosons. In \cite{Tasinato:2014eka} it was shown that  a decoupling 
limit exists in which the dynamics of the Goldstone bosons $\pi$ is described by Galileonic derivative
self-interactions.  
 This is a regime where some kind of equivalence theorem  should hold, with the physics  of the
 Goldstone bosons being equivalent to that of the longitudinal polarization of the vectors 
 (see for example \cite{Peskin:1995ev}). In our Higgs set-up, we can do one step
 further: we show that in this   decoupling limit, not only do the Goldstone self-interactions preserve Galileon invariance
 by themselves, but in addition they acquire new derivative couplings with the Higgs field $h$. These  automatically preserve the Galileon symmetry by assembling into bi-Galileon combinations.
  
  \smallskip
  
  To exhibit these features,
the limit
we have to consider is 
\be \label{limit1}
g\to 0\hskip0.5cm,\hskip0.5cm \lambda\to 0\hskip0.5cm,\hskip0.5cm \beta_{(8)}\to 0\hskip0.5cm,\hskip0.5cm v\to \infty\,\,,
\ee
such that 
\be \label{limit2}
m_A\to0\hskip0.5cm,\hskip0.5cm m_h\to0\hskip0.5cm,\hskip0.5cm \tilde \beta \to 0\hskip0.5cm,\hskip0.5cm
 \frac{\tilde \beta}{m_A^3}\,=\,{\rm fixed}\,\equiv \,\frac{1}{\Lambda_{g}^3}\,\,,
\ee
where $\Lambda_{g}$ is a mass scale that, as we will see in a moment, is associated with the strength of the Galileon interactions.
 Notice that the previous limits imply  that $g/m_{A}\,=\,1/v\,\to\,0$. 
 In order to have a correctly normalized kinetic term for the Goldstone boson $\pi$ we have to rescale this
 field,  and define $ \pi=\hat \pi/(\sqrt{2} m_A)$. Indeed the second term in the first line of  (\ref{explag}) becomes, in the limit (\ref{limit1}), 
 \bea
 -m_A^2 \left( A_\mu-\partial_\mu \pi\right)^2&=&- \left( m_A\,A_\mu-\frac{1}{\sqrt{2}}\partial_\mu \hat \pi\right)^2
 \nonumber\\
 &\to&-\frac12\,(\partial_\mu \hat \pi)^2\,,
 \eea
  so the Goldstone boson acquires a standard kinetic term.  In the limits (\ref{limit1},  \ref{limit2}), when expressed in terms
  of the canonically normalized Goldstone field $\hat \pi$, the total Lagrangian ${\cal L}_{tot}$ reduces to 
  \be
  {\cal L}_{tot}\,=\,-\frac14 F_{\mu \nu} F^{\mu \nu} -\frac12\,(\partial_\mu h)^2 -\frac12\,(\partial_\mu \hat \pi)^2 
  -\frac{1}{\Lambda_g^3} \,\left(\partial_\mu \hat \pi \partial^\mu \hat \pi\right) \,\Box \hat \pi-\frac{1}{3\,\Lambda_g^3}
  \left( \partial_\mu  h\, \partial^\mu  h \,\Box \hat \pi-
  \partial_\mu  h \,\partial^\nu  h \,\partial_\nu \partial^\mu \hat \pi
  \right)\,.
  \ee
Hence, as announced,  in this decoupling  limit the Lagrangian acquires a bi-Galileon structure, and the physical
Higgs itself acquires bi-Galileon couplings\footnote{The above bi-Galileon interaction corresponds to (\ref{eq:hhpi}) in Appendix \ref{app.B} but with $h$ and $\pi$ exchanged.} \cite{Deffayet:2010zh,Padilla:2010de} with the Goldstone boson describing the dynamics of the longitudinal
vector polarization. 
  The connection that we pointed out  with Galileons can help to render the structure of the theory stable under
   radiative corrections.  Galileon Lagrangians are known to enjoy powerful non-renormalization theorems \cite{Luty:2003vm,Nicolis:2004qq} 
   that might be 
    applied in the  present context to protect
     the size of the higher dimensional operators ${\cal L}_{(8),\,(12),\,(16)}$ that we introduced in this section. 
 We leave for future work the analysis of this point and move on to briefly discuss the possible phenomenological consequences and relevance of such interactions.

We can think to   two different ways 
   in which the Higgs field can couple to matter, that would 
   allow  to exploit the bi-Galileon interactions.
     The first is a  direct coupling of  the Higgs  $\phi$ to the trace of the energy momentum tensor $T$ via operators that respect gauge invariance such as for example $\phi^* \, \phi \,T$. In the case in which the Higgs scalar of our model is very light -- as might be required for cosmological applications --  such couplings could be associated with a long range force that needs to be screened. In our set-up we have  shown that, in an appropriate regime, the Higgs scalar combines with the longitudinal polarization of the vector to form bi-Galileon derivative combinations. These non-linear operators can then lead to a Vainshtein mechanism that is able to suppress the aforementioned long range force.

Other possible couplings involve derivative operators.
An  example among others is a  gauge invariant  coupling   of the form  
$({\cal D}_\mu \phi)^* ({\cal D}^\mu \phi) T$, where the ${\cal D}_\mu$ is a covariant derivative containing gauge fields (see eq \eqref{defcD}). 
  Once the covariant derivatives are expanded, such a combination leads among others to operators  of the form
   $A_\mu A^\mu T$,
    that couple vectors to the energy momentum tensor.
      More generally, one  could  generalize    the derivative disformal couplings of scalars to matter  proposed by Bekenstein \cite{Bekenstein:1992pj}, by promoting the standard derivative to covariant derivatives.
  It would be interesting to explore in detail the phenomenology of these derivative couplings. We can imagine that they could lead to long range forces, since the Higgs and the vector longitudinal polarization are  derivatively coupled to the energy momentum tensor. The bi-Galileon self-interactions discussed above can then provide the Vainshtein mechanism needed to screen them.

These arguments of course only scratch the surface of the possible couplings of our Higgs field to matter
and their phenomenological consequences. We hope to return to this subject with  a separate detailed publication.

\section{Higgs mechanism and generalized non-abelian symmetry breaking}
\label{sec-nonab}

The Higgs construction that we  developed in the abelian case can be directly  extended to
 the non-abelian case.
  This is interesting because, applying the St\"uckelberg approach in this case would be more
   laborious than in the abelian set-up.   
  Again we  focus
on theories that contain dimension-8 operators with  derivative self-interactions of the Higgs field.
 We investigate theories that spontaneously
break non-abelian symmetries, leading to consistent derivative self-interactions for gauge vectors,  and  generalizing the abelian symmetry
breaking case discussed in the previous section and in \cite{Tasinato:2014eka}. Instead of providing a fully general treatment,  
 we concentrate  on a representative example to make clear  our arguments.


We consider  an $SU(2)$ theory
with a doublet of complex scalars $\phi=\{\phi^\alpha\}$, with $\alpha=1,2$ transforming in the fundamental representation.
 The construction of a Higgs model for this theory, which spontaneously breaks the $SU(2)$ symmetry,  is a standard textbook  
  example, see e.g. \cite{Maggiore:2005qv}. Here we consider additional derivative self-interactions of the Higgs field, that lead to 
   derivative self-interactions of the gauge vectors. 
   
The  Lagrangian we are  interested in, is invariant under the non-abelian $SU(2)$ symmetry,  
 and is written,
\bea
{\cal L}_{SU(2)}=- \left( \cD^\mu \phi\right)^\dagger\,\cD_\mu \phi-V(\phi)
-\frac12\,\tr\left[ F_{\mu \nu} F^{\mu\nu} \right] +\,{\cal L}^{SU(2)}_{(8)}
\,.
\label{cubna1}
\eea
The field  $\phi$ is  our Higgs, that as stated above is a doublet under the $SU(2)$ symmetry; the covariant derivative acts on its components as
\be\label{nacd}
(\mc{D}_\mu \,\phi)^\alpha\,=\, \partial_\mu \phi^\alpha-i g A_\mu^a\,\left( T^a\right)^\alpha_{\,\,\beta}\,\phi^\beta
\,\,,
\ee
where $T^a$ are the generators in the fundamental representation, that for $SU(2)$ are 
proportional to the Pauli matrices, $T^a=\sigma^a/2$. 
The non-abelian transformation acts as 
\bea
\phi&\to& U \phi \,\,,
\\
A_\mu&\to& U A_\mu U^\dagger-\frac{i}{g}\,\left(\partial_\mu U \right) U^\dagger\,\,,
\eea
with $A_\mu\equiv A_\mu^a\,T^a$, and the transformation group element is $U\equiv \exp{\left[ i g \,\theta^a(x)\,T^a\right]}$. The covariant
derivative (\ref{nacd}) transforms as expected
\bea
(\cD_\mu \,\phi)&\to&U\,(\cD_\mu \,\phi)\,\,.
\eea
The field strength associated with the vector potential is defined as
\be
F_{\mu\nu}\,=\,\partial_\mu A_\nu-\partial_\nu A_\mu-i g\,\left[ A_\mu,\,A_\nu\right]
\,\,,
\ee
and transforms as 
\bea
F_{\mu\nu}&\to&U\,F_{\mu\nu}\,U^\dagger
\,\,,
\eea
the corresponding gauge invariant  
vector kinetic term is
\be
-\frac12\,\tr\left[ F_{\mu \nu} F^{\mu\nu} \right]\,=\,-\frac14\,F^a_{\mu \nu} F^{a\,\mu\nu} 
\,\,,
\ee
where we used the following identity valid for fundamental representations of the gauge group $\tr\left[T^a T^b\right]\,=\,
\frac12\,\delta^{a b}$. 
The Higgs potential is written as
\be
V(\phi)\,=\,\lambda\left( \phi\,\phi^\dagger-v^2\right)^2
\,\,,
\ee
and is invariant under the unitary transformations that we are considering. It is characterized
by  a family of degenerate vacua, with $\phi\,\phi^\dagger\,=\,v^2$, that spontaneously 
break the gauge symmetry. 

\smallskip

The dimension-8 Lagrangian 
${\cal L}^{SU(2)}_{(8)}$ in the second line of  (\ref{cubna1}), responsible
 for breaking the non-abelian symmetry in such a way to generate consistent derivative vector self-interactions, 
 is constructed similarly to what was done for the case of abelian symmetry in the previous section.
We define the gauge invariant tensor combinations
  \begin{align}
	L_{\mu\nu}&\equiv \frac12 \left
	[(\mc{D}_{\mu}\phi)^{\dagger}(\mc{D}_{\nu}\phi)+(\mc{D}_{\nu}\phi)^{\dagger}(\mc{D}_{\mu}\phi)\right]\,,
	 \\
	  \,Q_{\mu\nu} &\equiv \frac{i}{2} \left[
	  \phi \,\left( \mc{D}_{\mu}\mc{D}_{\nu}\phi\right)^\dagger
	  -\phi^\dagger\mc{D}_{\mu}\mc{D}_{\nu}\phi 
	  	  \right]\,,
\end{align}
built in terms of the Higgs doublet $\phi$. Then,
\bea
{\cal L}^{SU(2)}_{(8)}&\equiv&
-\frac{\beta}{\Lambda^4}
\,\left[
L_\rho^{\,\,\rho} Q_{\sigma}^{\,\,\sigma}-L_{\mu}^{\,\,\nu} Q_{\nu}^{\,\,\mu}
\right]\,\,,
\eea
with $\beta$ a dimensionless coupling constant, and $\Lambda$ a scale. 
For the very same arguments discussed in the abelian case, this dimension-8 operator is gauge invariant, 
and consistent since it does not introduce ghost degrees of freedom.

To proceed, we recall that $SU(2)$ transformations
are characterized by {\it three} free parameters, while our Higgs field has four independent real components. 
 At this stage,  
we can use the gauge freedom to 
fix a unitary  gauge and 
eliminate three of the Higgs  four components. We  write
\be
\phi\,=\,\binom{0}{v+\frac{1}{\sqrt 2}\,h}
\ee
with $\pi$ a real scalar field. 
 The covariant derivative acting on the Higgs becomes
\be
D_\mu \phi\,=\,\frac{1}{\sqrt{2}}\binom{0}{\partial_\mu h}-i\,\frac{g}{2}\,\left(v+\frac{1}{\sqrt2} h\right)
\,\binom{
A_\mu^{1}-
i A_\mu^{2}
}{-A_\mu^{3}}
\,.
\ee
On the other hand, the second covariant derivative  on the complex scalar $\phi$  acts as 
\bea
D_\nu D_\mu \phi&=& \partial_\nu \partial_\mu \phi^\alpha-i g \left(
 \partial_\nu
  A_\mu^c\right) \left(T^c\right)^\alpha_\gamma \phi^\gamma
  -i g 
  A_\mu^c  \left(T^c\right)^\alpha_\gamma \partial_\nu\phi^\gamma
   -i g 
  A_\nu^c  \left(T^c\right)^\alpha_\gamma \partial_\mu\phi^\gamma
  \nonumber
  \\
  &&
  -g^2 A_\nu^a A_\mu^b \, 
  \left(T^a\right)^\alpha_\beta \left(T^b\right)^\beta_\gamma\,\phi^\gamma\,\,.
\eea

\smallskip

Plugging these ingredients 
 in the expression (\ref{cubna1}) for  ${\cal L}_{SU(2)}$ and expanding, we
 find the following Lagrangian for the Higgs field $h$, the vectors $A_{\mu}^{a}$,
 and their couplings (sum over repeated indexes)  
 \bea
{\cal L}_{SU(2)}&=&
-\frac14\, F^{a}_{\mu \nu} F^{a\,\mu\nu} 
-\frac{g^2\,v^2}{4}
\,\left( A_\mu^a A^{a\,\mu}\right)
-\frac{\beta\,g^3\,v^4}{8\,\Lambda^4}\,\left[ \left(A^a_\mu\,A^{a\,\mu}  \right)\,\partial_\nu  A^{3\,\nu} 
-  \left(A^a_\mu\,A^{a}_{\nu}  \right)\,\partial^\mu  A^{3\,\nu} 
\right]
\nonumber
\\
&&
-\frac{1}{2} \partial_\mu h \partial^\mu h-2\,\lambda\,v^2\,h^2-\sqrt{2}\,\lambda\,v\,h^3-\frac{\lambda}{4}\,h^4
\nonumber
\\
&&
-\frac{\beta\,g\,v^2}{4 \,\Lambda^4}\,\left(\partial_\mu h \partial^\mu h\,\,\partial_\nu A^{3\,\nu}  -
\partial_\mu h \partial^\nu h\,\,\partial_\nu A^{3\,\mu} 
\right) \,\left( 1+\frac{\sqrt{2} h}{v}+\frac{h^2}{2\,v^2}\right)
\nonumber
\\
&&
-\frac{\beta\,g^3\,v^3}{4\,\sqrt{2} \,\Lambda^4}\,
\left(h+\frac{3 \,h^2}{2 \sqrt{2}\, v}+ \frac{h^3}{2\, v^2}
+\frac{h^4}{8 \sqrt{2}\,v^3}
\right)
\times
\nonumber
\\
&&
 \hskip0.5cm 
\times
\Big[\left(A^a_\mu\,A^{a\,\mu}  \right)\,\partial_\nu  A^{3\,\nu} 
 +
 A^{a\,\mu} A^{3\,\nu}\,\partial_\mu A_\nu^{a}  +A^a_\mu\,A^{3\,\mu} \,\partial_\nu  A^{a\,\nu}
\nonumber
\\
&& \hskip0.9cm 
 -  A^{a\,\mu} A^{a\,\nu}\,\partial_\mu A_\nu^{3} 
-2 \,A^{3\,\mu} A^{a\,\nu}\,\partial_\mu A_\nu^{a}   \Big]\,\,. \label{non-ab-lag}
\eea
Hence when the $vev$ $v\neq0$, this set-up
 spontaneosly breaks the non-abelian gauge symmetry.
   It  
 not only provides a mass to the three gauge bosons but also ghost-free derivative self-interactions among them that 
 corresponds to a non-abelian
 generalization of  \cite{Tasinato:2014eka}. 
   Moreover, it introduces new higher-dimensional couplings  (with or without derivatives) between the Higgs field 
   and the vector, proportional to the coupling constant $\beta$. 
 The Lagrangian  \eqref{non-ab-lag} is expressed in unitary gauge: if we were to re-introduce the would-be Goldstone bosons, we would find new interactions between them and the Higgs field, that in an appropriate decoupling
 limit leads to  a theory of multi-Galileons,  generalizing the findings of the previous section.

\section{Conclusions}
We presented a Higgs mechanism for spontaneously breaking a gauge symmetry,
  to obtain the non-linear derivative vector self-interactions recently studied in \cite{Tasinato:2014eka,Heisenberg:2014rta}, and extended the discussion
 to a case with non-abelian symmetry. 
 After symmetry breaking, the resulting theory contains the desired vector
 self-interactions, and in addition new ghost-free derivative interactions between the Higgs  and the vector bosons.  
 We studied some of the features of the resulting set-up. We
 showed that  the Lagrangian controlling the would-be Goldstone boson of this theory obtains a Galileon structure in an appropriate decoupling limit. Interestingly, in the same limit the would-be Goldstone boson also
   acquires     
 derivative couplings with the physical Higgs,  that combine in such a way to form a bi-Galileon system with fixed
 coefficients,  determined by gauge invariance. This suggests that, once we introduce an appropriate source, a Vainshtein mechanism should actively screen it from both the longitudinal mode of the vector and the Higgs field of the full theory.
  
  \smallskip
  
   Our results can be further  developed and extended, both from a
   phenomenological and a theoretical perspective. 
  From 
   the point of view of phenomenology, it 
    is known that vector theories with derivative self-interactions can have interesting cosmological applications \cite{Tasinato:2014eka,Tasinato:2014mia}.
    It 
   would be interesting to understand whether the new interactions associated with the  Higgs scalar can improve 
  the strong-coupling issues of  cosmological solutions driving  acceleration \cite{Tasinato:2014mia}, in a way
  resembling the quasi-dilaton extension of massive gravity \cite{D'Amico:2012zv}.  Namely, the inclusion of an additional degree of freedom together with its special structure suggests that the cosmology of our model could have a far richer phenomenology.  Also, it would be interesting to understand
  whether the Vainshtein-like screening mechanism that suppresses the effect of the longitudinal vector mode \cite{Tasinato:2014eka} is somehow modified by  the interactions with the Higgs scalar, possibly offering new suggestions for testing the theory.
  
The utility of the Vainshtein mechanism opens up the number of ways we can add couplings with other fields.  Without a screening mechanism, we would have to confine our theory to a dark sector such that there are no detectable interactions with the Standard Model.   At the end of Section \ref{sec-abelian} we briefly discussed how to couple the Higgs field to external matter.   It remains to be investigated whether our theory can be coupled to Standard Model fields in a gauge invariant way such that the interactions with the longitudinal mode of the vector and Higgs are screened.  This would necessarily entail addressing an open problem in the field.  That is, whether suitable Vainshtein mechanisms are possible beyond the very symmetrical and static matter distributions studied so far.  Specifically, we would like to consider whether currents formed from Standard Model particles, which usually source the electromagnetic field, could source in a non-linear way, the extra modes in the infrared sector of our theory.  In a different footing,  one needs  
  to understand whether our Higgs self-interactions can find some new applications 
  in particle physics model building, exploring the possibility that the Higgs
  field we discussed corresponds to the Standard Model Higgs.   
  
  From a more theoretical point of view, our set-up might  be regarded as a possible step towards UV-completions of theories closely related
  to Galileons.  Whether our Higgs construction can improve some of the high energy features 
  of the theory and have some role when studying quantum effects remains an open problem.  However we find it intriguing that our theory still exhibits a bi-Galileon 
  structure in an appropriate decoupling  limit, showing that the Higgs field does not ruin the Galileon symmetry. 
    It is also possible that  the structure of the higher dimensional Higgs operators
   we considered is somehow protected by non-renormalization theorems similar to the ones that
   apply to Galileon theories. 
   These observations might serve as a guide towards finding Higgs mechanisms for other theories related to Galileons, as for example massive gravity.

\acknowledgments
It is a pleasure to thank  
Dario Cannone, 
Javier Chagoya, Kevin Falls, Andrew Matas and Nicholas Ondo for useful discussions.  MH is support by a UK Science and Technology Facilities Council (STFC) research studentship.  KK is supported by the STFC grants ST/K00090/1 and ST/L005573/1. GT thanks the STFC for financial support through the grant ST/H005498/1.

\appendix

\section{Consistency of 
  our Higgs higher-dimensional interactions }
\label{app-ghostfree}

In this appendix we would like to develop some arguments  aimed 
 to show that the Higgs interactions contained in Lagrangians (\ref{defol}-\ref{defoq})
 are consistent, in the sense that they are free of
 ghost degrees of freedom.
 We specialize
 to
  the 
 case  of 
abelian symmetry breaking, but the same 
 arguments can be straightforwardly extended to the non-abelian case. 
The interactions in eqs.(\ref{defol}-\ref{defoq}) are built in terms of totally antisymmetric $\ve$-tensors. Once expanding the covariant derivatives acting on the Higgs field, and decomposing the Higgs in norm and phase as in the main text,  we find that there can arise three kinds of possibly dangerous combinations: 
 \bea
 \epsilon^{\alpha_1 \alpha_2\dots}\,\epsilon_{\beta_1  \beta_2 \dots} \partial_{\alpha_1} \partial^{\beta_1}
   \varphi\,\partial_{\alpha_2} \vphi  \partial^{\beta_2} \vphi\dots  \label{di1}
 \\
  \epsilon^{\alpha_1 \alpha_2\dots}\,\epsilon_{\beta_1  \beta_2 \dots} \, A_{\alpha_1}  A^{\beta_1} \,
  \,\partial_{\alpha_2} \vphi  \partial^{\beta_2} \vphi
  \dots  \label{di2}
  \\
    \epsilon^{\alpha_1 \alpha_2\dots}\,\epsilon_{\beta_1  \beta_2 \dots} \, \partial^{\beta_1} A_{\alpha_1}  \vphi \,\partial_{\alpha_2} \vphi  \partial^{\beta_2}\vphi \dots \label{di3}
   \eea
where the dots contain additional pieces, of the same type as the above, or other contributions that contain
single or no derivatives of $\varphi$ --  always contracted with the $\ve$-tensor.
 Interactions as the ones listed in (\ref{di1}-\ref{di3}), when appearing in the Lagrangian, 
are {\it a priori} dangerous  because they contain second derivatives acting on the scalar $\varphi$, and/or 
 the gauge potential $A_\mu$. We have to ensure that the corresponding equations of motion 
  do   not contain more than two space-time derivatives of 
   the fields involved. Moreover, the equation of motion for $A_0$ should not contain time derivatives acting on $A_0$
   itself, so to ensure that $A_0$ is a constraint. These requirements, together
   with the positivity of the kinetic terms,
   are sufficient to ensure the
    absence of ghosts.

 Interactions as (\ref{di1}) are the familiar scalar Galileon interactions \cite{Nicolis:2008in}: the structure of the $\ve$-tensors
 does not allow them to generate higher space-time derivatives in their equations of motion. Indeed, the equations
 of motion for a scalar field $\varphi$ can certainly  lead to derivatives acting on the first part, $ \partial_{\alpha_1} \partial^{\beta_1}\,\varphi$,   of (\ref{di1}) -- as for example contributions like  
 \be
 \epsilon^{\alpha_1 \alpha_2\dots}\,\epsilon_{\beta_1  \beta_2 \dots} \,
  \partial_{\alpha_2} \partial^{\beta_2}\,
 \partial_{\alpha_1} \partial^{\beta_1}\,\varphi\dots \hskip0.5 cm {\rm or}\hskip 0.5 cm
  \epsilon^{\alpha_1 \alpha_2\dots}\,\epsilon_{\beta_1  \beta_2 \dots} \,
  \partial_{\alpha_2}\,
 \partial_{\alpha_1} \partial^{\beta_1}\,\varphi\dots
 \ee
But the  $\ve$-tensor makes them vanishing: the operator $ \partial_{\alpha_1}\,
 \partial_{\alpha_2}$ is symmetric on its indexes, and gives zero when contracted with the $ \epsilon^{\alpha_1 \alpha_2\dots}$. This fact is familiar and was developed in \cite{Deffayet:2009mn}. Similar arguments can be made to show
 that (\ref{di2}), (\ref{di3}) cannot contribute to the equation
  of motion for $A_0$ with terms containing the time derivative of $A_0$ itself (see also \cite{Tasinato:2014eka,Heisenberg:2014rta}).
   Since $A_\mu$ is always contracted with the $\ve$-tensor, it is simple to convince
    oneself 
    that the only possibly dangerous contributions from the equation of motion of $A_0$ -- that is the ones
    that might have time derivatives acting on $A_0$ -- 
      are pieces that contain first or second derivatives 
    acting on the gauge potential, as
 \be
 \epsilon^{0\, \dots}\,\epsilon_{\beta_1  \beta_2 \dots} \, \partial^{\beta_1} A^{\beta_2}\,,
 \hskip0.7cm {\rm or} \hskip0.7cm
  \epsilon^{0\, \alpha_2\dots}\,\epsilon_{\beta_1   \dots} \, \partial^{\beta_1} A_{\alpha_2}\,,
 \hskip0.7cm {\rm or} \hskip0.7cm
  \epsilon^{0\, \alpha_2\dots}\,\epsilon_{\beta_1  \beta_2 \dots} \, \partial_{\alpha_2} \partial^{\beta_1} A^{\beta_2}  
 \,.
 \ee
 In the first option, the index  $\beta_1$ and $\beta_2$ can not simultaneously take the value zero, due to the antisymmetric property of
  the $\ve$-tensor, hence this contribution vanishes for the possibly dangerous case.  A similar argument exists for the second 
  and third option. The crucial fact is that one of the indexes of the $\ve$-tensor is already fixed to be zero since
  we are 
  evaluating the equation of motion for $A_0$;  
  hence, $\alpha_2\neq0$  and we cannot have time derivatives acting on $A_0$.

\section{Ghost free scalar-vector interactions} \label{app.B}
	
\subsection{bi-Galileons}
	We wish to find ghost free derivative couplings between a scalar $\pi$ and a vector field $A_{\mu}$.  In order to achieve this, we will find it useful to first consider `bi-Galileon' interactions.  Bi-Galileons are an extension to two scalar fields of the original Galileon theory first introduced to cosmology in \cite{Nicolis:2008in}.  (Their properties, however, were first discussed in \cite{Fairlie:1991qe} for a rather different purpose.)  \\A Galileon is a scalar field $\pi$ the action of which is invariant under Galilean shifts in its field space, $\pi \rightarrow \pi + b_{\mu}x^{\mu}+c$.  They have the property that although their actions contain both first and second order derivatives, their equations of motion are of second order only.  Furthermore, it was shown in \cite{Nicolis:2008in,Fairlie:1991qe} that, up to total derivatives, there is a unique term for each order in the field $\pi$ up to $n+1$, where $n$ is the dimension of the space-time.  This is due to the fact that each nontrivial derivative term is associated with one Cayley invariant of the matrix $\p_{\mu}\p_{\nu}\pi$.\\ 
	We make use of the $\mathit{Levi-Civita}$ epsilon tensor to write the Lagrangian for the Galileons in a compact form \cite{Deffayet:2009mn}.  Using the following property:
		\be
			\ve_{\gam_1 \dots \gam_{D-n} \al_{1}\dots \al_{n}}\ve^{\gam_1 \dots \gam_{D-n} \beta_{1}\dots \beta_n} = -(D-n)!\,n!\,\delta^{[\beta_{1}\dots \beta_n]}_{\al_{1}\dots\al_n}
		\ee
		where the square brackets represent normalised anti-symmetric permutations, we can write the Galileon Lagrangians as:
		\begin{align}
			\mc{L}_1 &= \pi \\ \mc{L}_2 &= \frac{1}{3!}\ve^{\mu_1\nu \lam \gam} \ve^{\mu_2}_{\,\,\,\,\,\,\nu \lam \gam} \pi_{\mu_1}\pi_{\mu_2} := \mc{E}_{(2)}\pi_1\pi_2 \\ \mc{L}_3 & = \frac{1}{2!}\ve^{\mu_1 \mu_3 \nu \lam}\ve^{\mu_2 \mu_4}_{\,\,\,\,\,\,\,\,\,\,\,\, \nu \lam} \pi_{\mu_1}\pi_{\mu_2}(\pi_{\mu_3\mu_4}) := \mc{E}_{(4)}\pi_1 \pi_2 (\pi_{34}) \\ \mc{L}_4 & =  \ve^{\mu_1 \mu_3 \mu_5 \nu} \ve^{\mu_2 \mu_4 \mu_6}_{\,\,\,\,\,\,\,\,\,\,\,\,\,\,\,\,\,\,\, \nu} \pi_{\mu_1}\pi_{\mu_2}(\pi_{\mu_3\mu_4}\pi_{\mu_5 \mu_6}) := \mc{E}_{(6)}\pi_1 \pi_2 (\pi_{34}\pi_{56}) \\ \mc{L}_5 & =  \ve^{\mu_1 \mu_3 \mu_5 \mu_7} \ve^{\mu_2 \mu_4 \mu_6 \mu_8}\pi_{\mu_1}\pi_{\mu_2}(\pi_{\mu_3\mu_4}\pi_{\mu_5 \mu_6}\pi_{\mu_7 \mu_8}) := \mc{E}_{(8)}\pi_1 \pi_2 (\pi_{34}\pi_{56}\pi_{78})
		\end{align} 
		Where we have defined $\mathcal{E}^{1234 \dots}_{2n} = \frac{1}{(D-n)!}\ve^{135 \ldots \nu_1\nu_2 \ldots \nu_{D-n}}\ve^{246 \ldots}_{\,\,\,\,\,\,\,\,\,\,\,\,\,\,\,\nu_1\nu_2 \ldots \nu_{D-n}}$ which has been written in short hand as $ \mc{E}_{(2n)}$ and the numbers are short hand for labeled indices: $\{\mu_1 \mu_2 \dots\}$. Furthermore, we have that $\pi_{\mu_1\dots\mu_n} \equiv \p_{\mu_n}\dots \p_{\mu_1}\pi$.\\ 
		With this notation it is very easy to see that the variation of these Lagrangians would never have higher than two derivatives. For instance, taking the variation of $\mc{L}_5$ gives us: 
		\begin{align}
			0 =& \,\del \mc{S}_5 = \int \mathrm{d}^4 x \,\del \mc{L}_5 \notag \\ =& \int \mathrm{d}^4 x \,\mc{E}_{(8)}\Big[ 2\del\pi_1\pi_2(\pi_{34}\pi_{56}\pi_{78}) + 3\pi_1\pi_2 (\del\pi_{34}\pi_{56}\pi_{78})\Big] \notag \\ =& \int \mathrm{d}^4 x \, \mc{E}_{(8)} \Big[-2\p_1\big(\pi_{2}\pi_{34}\pi_{56}\pi_{78}\big)-3\p_3\p_4\big(\pi \pi_{12} \pi_{56}\pi_{78}\big)\Big]\del\pi \notag \\ =& -5 \int \mathrm{d}^4 x \,  \mc{E}_{(8)}(\pi_{12}\pi_{34}\pi_{56}\pi_{78})
		\end{align}
		Where we have integrated by parts and found that the only term to survive the summation with the totally antisymmetric tensor $\mc{E}_{(8)}$ has, indeed, only derivatives of second order.\\
	
	Bi-Galileons were first introduced in a general setting in \cite{Deffayet:2010zh} and were treated in depth in \cite{Padilla:2010de}.  The action for the two scalar fields $\pi$ and $h$, is invariant under separate Galilean transformations: $\pi \rightarrow \pi + b^{(\pi)}_{\mu}x^{\mu}+c^{(\pi)}$ and $h \rightarrow h+ b^{(h)}_{\mu}x^{\mu}+c^{(h)}$.  Furthermore, the equations of motion for both fields are exactly second order in their derivatives.  We use the notation introduced above and follow the methods outlined in \cite{Deffayet:2010zh}.\\
	First we enforce a symmetry relation.  That is, $\mathcal{L}_{h\pi} = \mathcal{L}_{\pi h}$ with $h \leftrightarrow \pi$.  I.e.
		\be
			\ve^{\mu\nu\rho\lam}\ve_{\al\beta\gam\lam}\pi_{\mu}h^{\al}(h_{\nu}^{\,\beta}h_{\rho}^{\,\gam}) \rightarrow \ve^{\mu\nu\rho\lam}\ve_{\al\beta\gam\lam}h_{\mu}\pi^{\al}(\pi_{\nu}^{\,\beta}\pi_{\rho}^{\,\gam}) 			
		\ee
			
	It will be important to remember this choice when we substitute the vector for one of the Galileons.\\
	The general Lagrangian can be written as the sum of the following sub-Lagrangians:
			
		\begin{itemize}
			
			\item[$\mathcal{E}_{(8)}$:]
			\begin{flalign} \al_{(5,0)} \mathcal{L}_{(5,0)}&= \al_{(5,0)} \mathcal{E}_{(8)}h_{1}h_{2}(h_{34}h_{56}h_{78})& \\ \al_{(4,1)}\mathcal{L}_{(4,1)}&= \al_{(4,1)} \mathcal{E}_{(8)}h_{1}\pi_{2}(h_{34}h_{56}h_{78})&  \\ \al_{(3,2)} \mathcal{L}_{(3,2)}&= \al_{(3,2)} \mathcal{E}_{(8)}h_{1}\pi_{2}(\pi_{34}h_{56}h_{78})&
			\end{flalign}
				
			\item[$\mathcal{E}_{(6)}$:]
			\begin{flalign}
			\al_{(4,0)} \mathcal{L}_{(4,0)}&= \al_{(4,0)} \mathcal{E}_{(6)}h_{1}h_{2}(h_{34}h_{56})& \\ \al_{(3,1)} \mathcal{L}_{(3,1)}&= \al_{(3,1)} \mathcal{E}_{(6)}h_{1}\pi_{2}(h_{34}h_{56})& \\ \al_{(2,2)} \mathcal{L}_{(2,2)}&= \al_{(2,2)} \mathcal{E}_{(6)}h_{1}\pi_{2}(\pi_{34}h_{56})&
			\end{flalign}
				
			\item[$\mathcal{E}_{(4)}$:]
			\begin{flalign}
			\al_{(3,0)} \mathcal{L}_{(3,0)}&= \al_{(3,0)} \mathcal{E}_{(4)}h_{1}h_{2}(h_{34})& \\ \al_{(2,1)} \mathcal{L}_{(2,1)}&= \al_{(2,1)} \mathcal{E}_{(4)}h_{1}\pi_{2}(h_{34})\label{eq:hhpi}&
			\end{flalign}
			
			\item[$\mathcal{E}_{(2)}$:]
			\begin{flalign}
			\al_{(2,0)} \mathcal{L}_{(2,0)}&= \al_{(2,0)} \mathcal{E}_{(2)}h_{1}h_{2}& \\ \al_{(1,1)} \mathcal{L}_{(1,1)}&= \al_{(1,1)} \mathcal{E}_{(2)}h_{1}\pi_{2}&
			\end{flalign}
				
			\item[$\mathcal{E}_{(0)}$:] 
			\begin{flalign}
			\al_{(1,0)} \mathcal{L}_{(1,0)}&= \al_{(1,0)} \mathcal{E}_{(0)}h&
			\end{flalign}
				
		\end{itemize}
	Where for each sub-Lagrangian we have the corresponding symmetrical exchange of the two fields: $ \beta_{(m,n)}\mathcal{L}_{(m,n)} = \beta_{(m,n)} \mathcal{E}_{(2(m+n-1))}\pi_1 h_2(\pi_{34}\ldots)$.\\
		
	\subsection{Bi-vectors and the scalar-vector Lagrangian}
	The above bi-Galileon terms can be identified as the decoupling limit of an interaction between a scalar and a vector.  Due to their special properties, these interactions cannot induce a ghostly fourth mode, (ghost free scalar-vector interactions were discussed in a different context in \cite{Khosravi:2014mua}).  We construct these interaction terms by first considering the products of two vectors, $X_{\mu}=\{A_{\mu},B_{\mu}\}$ with their derivatives, $X_{\mu\nu} \equiv \p_{\mu}X_{\nu}=\{\p_{\mu}A_{\nu},\p_{\mu}B_{\nu}\}$ and then substituting $B_{\mu}\equiv \p_{\mu}h$:
		\be
			\mc{L}_{\textrm{bi-vector}}= \mc{E}_{(2n)}X_1X_2(X_{\{34\}}\dots X_{\{2n-1\,2n\}})
		\ee
	Where we use $\{\,\,\} := (\,\,)\,\textrm{or} \,[\,\,]$ to indicate symmetric and anti-symmetric combinations respectfully.\\  
	When we constructed the Galileons above we relied on the fact that the indices associated with the partial derivatives acting on the scalar field commute (i.e. $\pi_{\mu\nu}=\pi_{\nu\mu}$).  For vectors, however, this is not true as the indices associated with the vector cannot be commuted (anti-commuted) with the indices associated with the partial derivative (i.e. $\p_{\mu}A_{\nu} \ne \p_{\nu}A_{\mu}$) and thus we need to take into account the new combinations that are possible.  This subtlety was discussed for a single gauge field in \cite{Heisenberg:2014rta} where it was found that one extra parameter is needed for both the quartic and quintic vector Galileons.  Furthermore, notice that, in the decoupling limit, these separate scalar-vector interactions converge to the same bi-Galileon term as they are related by the symmetry outlined above.  Although we add some redundancy due to some terms differing only by a total derivative, it is convenient to construct our Lagrangian by choosing $X_n:=  aA_n + bB_n$ and $X_{\{nm\}} := a A_{\{nm\}} + b B_{\{nm\}}$.\\  In order to make contact with the main text, in the following we consider only terms up to cubic order in the fields.  In such a case, we find that the terms with $X_{[nm]}$ cancel and we have:
		\begin{align}
			\mc{L}^{(3)}_{\textrm{bi-vector}}&= \mc{E}_{(4)}X_1 X_2 (X_{\{(34)\}})\notag \\ &=  \mc{E}_{(4)}(aA_1+bB_1)(aA_2 + bB_2)(aA_{(34)}+bB_{(34)})\notag \\ &= \mc{E}_{(4)}\Big\{ a^3 A_1 A_2(A_{(34)}) +a^2 b[A_1 A_2 (B_{(34)}) + 2 A_1 B_2(A_{(34)}) ]\notag \\&\qquad \qquad+ \mathrm{exchange}\, \{aA_n, aA_{(nm)}\}\longleftrightarrow \{bB_n, bB_{(nm)}\}\Big\}
		\end{align}
	Substituting $\p_{\mu}h$ for $B_{\mu}$ gives us the cubic scalar-vector interactions:		
			\begin{align}
			\al_{(3,0)} \mathcal{L}^{sv}_{(3,0)}&= \al_{(3,0)} \mathcal{E}_{(4)}A_{1}A_{2}(A_{(34)})  \\ \al_{(2,1)} \mathcal{L}^{sv}_{(2,1)}&= \al_{(2,1)} \mathcal{E}_{(4)}A_{1}A_{2}(h_{34}) \\ \al_{(2,1)'} \mathcal{L}^{sv}_{(2,1)'}&= \al_{(2,1)'} \mathcal{E}_{(4)}A_{1}h_{2}(A_{(34)})\\ \beta_{(0,3)} \mathcal{L}^{sv}_{(0,3)}&= \beta_{(0,3)} \mathcal{E}_{(4)}h_{1}h_{2}(h_{34}) \\ \beta_{(1,2)} \mathcal{L}^{sv}_{(1,2)}&= \beta_{(1,2)} \mathcal{E}_{(4)}h_{1}h_{2}(A_{(34)})\\ \beta_{(1,2)'} \mathcal{L}^{sv}_{(1,2)'}&= \beta_{(1,2)'} \mathcal{E}_{(4)}h_{1}A_{2}(h_{34})
			\end{align}			
	Where, 
	\be
    	\begin{cases}
      	 \al_{(n,m)} = a^n b^m \, \textrm{and} \, \al_{(n,m)'} = 2 a^n b^m & \text{if } n>m,\\
      	\beta_{(n,m)} = a^n b^m \, \textrm{and} \, \beta_{(n,m)'} = 2a^n b^m & \text{if } n<m.
   	 \end{cases}
	\ee
	These interactions return to the above cubic bi-Galileon terms in the appropriate decoupling limit, or rather, under substituting $A_{\mu}$ with $\p_{\mu}\pi$.  Lastly, for the convenience of the reader, we expand below the terms which correspond to the interactions generated in the main text:
	\begin{flalign}
			\al_{(3,0)} \mathcal{L}^{sv}_{(3,0)}&= \al_{(3,0)} \mathcal{E}_{(4)}A_{1}A_{2}(A_{34}) = - \al_{(3,0)}2!\delta^{[\mu_2\mu_4]}_{\mu_1 \mu_3} A^{\mu_1}A_{\mu_2}(A^{\mu_3}_{\,\,\mu_4})& \notag \\& = -\al_{(3,0)} 2 \,A^2 (\p \cdot A)+ \al_{(3,0)} \frac{1}{2}A_{\mu}(\p^{\mu}A^{\nu}+\p^{\nu}A^{\mu})A_{\nu}& \notag \\&= -\frac{5}{2}\al_{(3,0)}\,A^2(\p \cdot A)& \\ \beta_{(1,2)} \mathcal{L}^{sv}_{(1,2)}&= \beta_{(1,2)} \mathcal{E}_{(4)}h_{1}A_{2}(h_{34}) = - \beta_{(1,2)} 2! \delta^{[\mu_2 \mu_4]}_{\mu_1 \mu_3} h^{\mu_1}A_{\mu_2}(h^{\mu_3}_{\,\,\mu_4})& \notag \\ &= - \beta_{(1,2)} (\p^{\mu}h A_{\mu}(\Box h)-\p^{\mu}h A^{\nu}(\p_{\mu}\p_{\nu}h))&\notag \\&=- \beta_{(1,2)} (\p_{\mu}h A^{\mu}(\Box h)+\p^{\nu}\p_{\mu}h A_{\nu}\p^{\mu}h + \p_{\mu}h\p^{\nu}A_{\nu}\p^{\mu}h)&\notag \\&= - \beta_{(1,2)} ( (\p h)^2 \p\cdot A - \p_{\nu}h\p^{\nu}A^{\mu}\p_{\mu}h)&\label{cubexp}
	\end{flalign}



\end{document}